\documentclass{article}
\usepackage{graphicx}
\usepackage{multirow}
\usepackage{hyperref}

\begin{document}

\title{Testing Quantum Gravity}
\author{Johan Hansson\footnote{c.johan.hansson@ltu.se} \, \& Stephane Francois\\
 \textit{Division of Physics} \\ \textit{Lule\aa \,University of Technology}
 \\ \textit{SE-971 87 Lule\aa, Sweden}}

\date{}

\maketitle

\begin{abstract}
The search for a theory of quantum gravity is the most fundamental problem in all of theoretical physics, but there are as yet no experimental results at all to guide this endeavor. What seems to be needed is a pragmatic way to test if gravitation really occurs between quantum objects or not. In this article we suggest such a potential way out of this deadlock,
utilizing macroscopic quantum systems; superfluid helium, gaseous Bose-Einstein condensates and ``macroscopic" molecules.
It turns out that true quantum gravity effects - here defined as observable gravitational interactions between truly quantum objects - could and should be seen (if they occur in nature)
using existing technology. A falsification of the low-energy limit, in the accessible weak-field regime, would also falsify the full theory of quantum gravity, making it enter the realm of testable, potentially falsifiable theories, \textit{i.e.}
becoming real physics after almost a century of pure theorizing. If weak-field gravity between quantum objects is shown to be absent (in the regime where the approximation should apply), we know that gravity then is a strictly classical phenomenon absent at the quantum level.
\end{abstract}
The ``holy grail" of fundamental theoretical physics is quantum gravity - the goal of somehow reconciling gravity with the requirement of formulating it as a quantum theory, \textit{i.e.} ``explaining" how gravity as we presently know it emerges from some more fundamental microscopic theory. The most serious obstacle - from the point of view that physics is supposed to be a natural science telling us something about the real world - is the total lack of experiments guiding us. Today there are as yet no detected observational or experimental signatures of any quantum gravitational effects. Naively, essentially from pure dimensional analysis arguments, quantum gravity experimentally seems to require an energy of roughly $E_P = \sqrt{\hbar c^5 /G} \simeq 10^{28}$ eV, the ``Planck Energy" (or equivalently, the means for exploring length-scales of the order of the ``Planck Length", $l_P = \sqrt{\hbar G/c^3} \simeq 10^{-35}$ meters). Using existing technology, this would require a particle accelerator larger than our galaxy - so direct tests of quantum gravity seems, at first sight, impossible.

However, as quantum theory is supposed to be universal - no maximum length built into its domain of applicability - a \textit{low-energy}, large length-scale, formulation of the theory should still apply. A falsification of the low-energy limit, in the experimentally accessible weak-field regime, would also falsify the full theory of quantized gravity \cite{Hansson}, hence making it possible to test, and potentially rule out, quantum gravity with existing or near-future technologies\footnote{``Now if I consider only gravitostatics, I still have a problem. I still have a quantum theory of gravity.", R.P. Feynman \cite{Feynman1}.}. In fact, direct tests of the high-energy limit of general quantum gravity may never be possible. In that case high-precision laboratory tests of weak-field quantum gravity will be the only possibility to make quantum gravity a physical (testable/falsifiable) theory instead of merely a mathematical one (as it has been until now).

But how can a quantum theory be applied to the fairly large bodies needed?\footnote{Previous work purporting to having seen quantum gravity effects have in reality only probed the ``correspondence limit" of extremely high excitation \cite{Hansson}, in the classical gravitational field of the whole earth, \textit{e.g.} \cite{COW}, \cite{Nezvishevsky}.} The answer lies in macroscopic systems still obeying the rules and laws of quantum theory - in essence those described by macroscopic wavefunctions. For a free-falling, effectively two-body problem, it should then in principle be possible to measure, \textit{e.g.}, the resulting quantum gravitational excitation energies \cite{Hansson}. A positive result would show that the gravitational field is quantized, just like the quantized energy levels resulting from the Schr\"{o}dinger equation for hydrogen is implicit proof of the quantization of the electromagnetic field. We can immediately think of four such candidates (and combinations of them, and more fundamental electrically neutral particles like neutrons $\sim 10^{-27}$ kg):

i) Superfluid helium-II.

ii) Gaseous Bose-Einstein condensates ($\leq 10^9$ u $\sim 10^{-17}$ kg, presently).

iii) Buckyballs or other ``macroscopic" molecules known to still obey quantum mechanics ($\leq 10^4$ u $\sim 10^{-22}$ kg, presently).

iv) Neutron stars, believed to contain a substantial portion of their mass as superfluid neutrons \cite{NS}, which should give very significant quantum gravity effects, for instance potentially measurable as \textit{quantized} (discrete) gravitational redshift, the normal component acting incoherently (where each neutron interacts individually with the test particle - adding probabilities not amplitudes), not screening the effect.

For superfluids, as the temperature decreases below the $\lambda$-transition the superfluid component rapidly approaches 100\%. The helium atoms then condense into the same lowest energy quantum ``groundstate" (losing their individual identities) and it becomes \textit{the} state of the macroscopic superfluid. Hence, the superfluid is described by a single quantum wavefunction, even though macroscopic in size and mass \cite{Feynman}, and the same applies for gaseous Bose-Einstein condensates. It can then only behave in a completely ordered way, in which the action of any atom is correlated with the action of all the others, and thus has extreme sensitivity to ultraweak forces (like gravity).

So, if superfluid systems, dominated by the superfluid state, interact solely/mainly through gravity with other quantum systems, we can obtain a test of low-energy quantum gravity. As the whole quantum ``object" is described by a single wavefunction, quantum gravity affects, and is affected by, its whole mass.

We may consider several such possibilities:

A superfluid ($M$) gravitationally binding a mass ($m$) of either a) a neutral quantum particle such as a neutron, b) an atomic Bose-Einstein condensate or c) a ``macroscopic" quantum molecule. The system being in free-fall, inside a spherical Faraday cage, either in an evacuated drop-tower experiment on earth, in parabolic flight, or, ultimately, in permanent free-fall in a satellite experiment, \textit{e.g.} at the International Space Station, or a dedicated satellite similar to the European Space Agency ``STE-Quest" space mission proposal (Space-Time Explorer and QUantum Equivalence principle Space Test).

Also, a neutron star ($M$) plus "test-particle" ($m$) should exhibit substantial quantum gravity effects. Unfortunately,
the formalism in \cite{Hansson} is strictly applicable only to weak fields where the static (potential) gravitational contribution overwhelms the dynamical.

However, just like newtonian gravity is the weak-field/low-energy limit of general relativity, newtonian quantum gravity must be the weak-field/low-energy limit of general (presently unknown) quantum gravity. The main advantage being that newtonian quantum gravity is known and well-defined, and hence, in principle, testable \textit{today}. If weak-field gravity between quantum objects is falsified (in the regime where it should apply), we know that general quantum gravity is falsified too, meaning that gravity is then a strictly classical phenomenon absent at the quantum level.

The gravitational energy levels between quantum systems $m$ and $M$ are \cite{Hansson}
\begin{equation}
E_n (grav) =  - \frac{G^2 \mu m^2 M^2}{2 \hbar^2}\frac{1}{n^2} = - E_g \frac{1}{n^2}, (n = 1,2,3,...),
\end{equation}
where
\begin{equation}
\mu = \frac{mM}{m+M},
\end{equation}
is the reduced mass, introduced to facilitate any combination of masses ($\mu$ giving just $m$ for $m \ll M$, and $\mu = m/2$ if $m=M$), and
\begin{equation}
E_g = \frac{G^2 \mu m^2 M^2}{2 \hbar^2},
\end{equation}
is the quantum gravitational binding energy, \textit{i.e.} the energy required to
totally free the mass $m$ from $M$ in analogy to the Hydrogen
case, whereas the most probable radial distance is
\begin{equation}
\tilde{r}_{grav} \simeq \frac{n^2 \hbar^2}{G
\mu m M}.
\end{equation}
All analytical solutions to the normal Schr\"{o}dinger equation, the
hydrogen wavefunctions, carry over to the gravitational case with
the simple substitution $e^2/4 \pi \epsilon_0 \rightarrow GmM$, which is equivalent to replacing the reduced Bohr-radius, $a^*_0$, with the reduced ``gravitational Bohr-radius" \cite{Hansson}
\begin{equation}
b^*_0  = \frac{\hbar^2}{G\mu mM}
\end{equation}
in the wavefunctions
\begin{equation}
\psi_{nlm}  = R(r) \Theta(\theta) \Phi(\phi) = N_{nlm} R_{nl}
Y_{lm}.
\end{equation}
Here $N_{nlm}$ is the normalization constant, $R_{nl}$ the radial
wavefunction, and $Y_{lm}$ the spherical harmonics containing the
angular parts of the wavefunction. The gravitational Bohr-radius, $b^*_0$, also gives the distance where the probability density of the ground state $\psi_{100}$ peaks (and also the innermost allowed radius of orbits in the old semi-classical Bohr-model, equivalently, the radius where the circumference $2 \pi r$ equals exactly one deBroglie wavelength).

If we introduce the Planck mass
\begin{equation}
m_P  = \sqrt{\hbar c/G} \simeq 2.2 \times 10^{-8} \, kg,
\end{equation}
conventionally believed to be fundamental in quantum gravity, we can rewrite the quantum gravitational binding energy and the reduced gravitational Bohr radius as
\begin{equation}
E_g = \frac{\mu c^2}{2}\frac{m^2 M^2}{m^4_P},
\end{equation}
\begin{equation}
b^*_0  = \frac{\hbar}{\mu c}\frac{m^2_P}{m M},
\end{equation}
where $\hbar/\mu c$ in the last equation is just the reduced Compton wavelength for $\mu$. With $m$ = $M$ = $m_P$ this yields $E_g = E_P /4$, \textit{i.e.} 1/4th the Planck energy, and $b^*_0 = 2 l_P$, twice the Planck length, consistent with the naive expectation.

The quantum gravitational energy-levels of the system are as quoted above \cite{Hansson}. For example, for a mass $m=M=8.6 \times 10^{-14}$ kg the first few excited states above the groundstate would require $E_{1-2}= 2.2$ eV, $E_{1-3}= 2.6$ eV, $E_{1-4}= 2.8$ eV.

One possibility (but by no means the only one) to investigate ``quantum jumps" between these gravitational quantum states, and hence potentially detect the quantization of the gravitational field, would be to use a laser calibrated to these energy frequencies to experimentally detect and manipulate them. The system should not ``jump" until the laser is in resonance with the possible quantum gravitational states of the system. It should be noted that the excitation of the states are then effected by electromagnetism, whereas the decay towards the ground state would be gravitational transitions with graviton emission. Even if the gravitational decay is incredibly slow/improbable (depending on the combinations of $m$ and $M$) it is sufficient to observe photon absorption at the predicted resonance frequencies to verify the quantum gravity effect. (This being analogous to the fast production of \textit{e.g.} strange particles, via the strong interaction, and their subsequent slow decay via the weak interaction.)
An absorption spectrum will thus give the ``fingerprint" of quantum gravity in the system under consideration. If the masses could be chosen to give well separated energy-states in the energy range of visible light ($1.7$ eV $< E < 3.2$ eV), this would be completely analogous to optical absorption spectra in cold gases. As it nowadays is possible to identify single quanta with essentially 100\% efficiency, having just one system (instead of billions of atoms in gases) should not be an impossible obstacle in principle. For ease of visualization and analogy with familiar physics we have so far mentioned visible light. As seen in Table 1, and Figures 1 \& 2, maser energies hold more promise. Still, it turns out that it is rather hard to find the ``sweet-spot" where both $E_g$ and $b^*_0$ simultaneously are physically reasonable and potentially measurable. Fortunately, one can, however, tailor $m$ and $M$ so as to avoid coinciding with naturally occurring electromagnetic (\textit{i.e.} not quantum gravitational) spectral lines, in principle giving a unique ``smoking-gun" signal for quantum gravity.

An independent, qualitative argument indirectly implying the existence of quantum gravity - assuming the equivalence principle holds for rotating superfluid helium - is the effect in an annular ``torus-shaped" container of radius $R$ and annular width $d \ll R$. The frequency of rotation is then quantized, and consequently the energy of rotation is
\begin{equation}
E_j  = j^2 \frac{\hbar^2}{2 m R^2},
\end{equation}
where $j=(0,1,2,...)$. For $m = m_{^4He} \simeq 6 \times 10^{-27}$ kg, and $R \simeq 10^{-3}$ m, $
\hbar^2/2 m R^2 \simeq 5 \times 10^{-18}$ eV.

According to the equivalence principle, the physical basis of general relativity, gravitation is equivalent to acceleration, which in this case is
\begin{equation}
a  = j^2 \frac{\hbar^2}{m^2 R^3},
\end{equation}
and as the acceleration is quantized, so is the equivalent gravitation.
For the same parameter-values as above
$\hbar^2/m^2 R^3 \simeq 3 \times 10^{-7}$  m/s$^{2}$.

However, we immediately see that the groundstate ($j=0$) does not accelerate at all, \textit{i.e.} the equivalent quantum gravitational groundstate is unaffected and cannot ``fall", just like an electron cannot fall into the nucleus of an atom, which may resolve singularity problems arising in the classical theory. (Giving an innermost allowed gravitational ``orbit" in the old interpretation of Bohr, its circumference being exactly one deBroglie wavelength, while $\hbar \rightarrow 0$ in Eqs. (3) and (5) gives back the classical singularity, averted by the quantum condition $\hbar \neq 0$ really valid in nature.)

In a simply connected vessel (no ``hole") the total angular momentum is still quantized, but there can no longer be any bulk rotation as the superfluid is irrotational (the hole in the torus being what allows this in such non-simply connected vessels). Below the first critical angular velocity the superfluid is stationary. As the circulation reaches $\kappa = h/m \simeq 10^{-7}$ m$^2$/s a first quantum vortex will form, at  $2h/m$ a second one will appear, and so on. The resulting quantum vortices, $N$ individual ones all with $j=1$ as higher $j$ are unfavourable energetically \cite{Feynman}, should also be directly related to quantum gravity through the equivalence principle. As the core of the quantum vortex is of the order $R \sim$ 1{\AA}, the energy and acceleration for a single ``fundamental" vortex is $E_1 \sim  10^{-4}$ eV and $a \sim 10^{14}$ m/s$^{2}$. The non-rotating groundstate has no circulation, so no acceleration and again no equivalent effective gravity.

In conclusion, we have seen how quantum gravity in principle can be tested today, \textit{e.g.} using the quantum gravitational behavior of combinations of macroscopic superfluids, large molecules, Bose-Einstein condensates and neutrons. Indirectly, the observed quantized rotation/acceleration of superfluids already hints at the existence of quantum gravity. However, this assumes that the equivalence principle is still valid at the quantum level, which is far from proven.
\begin{figure}
\begin{center}
\scalebox{1}{\includegraphics {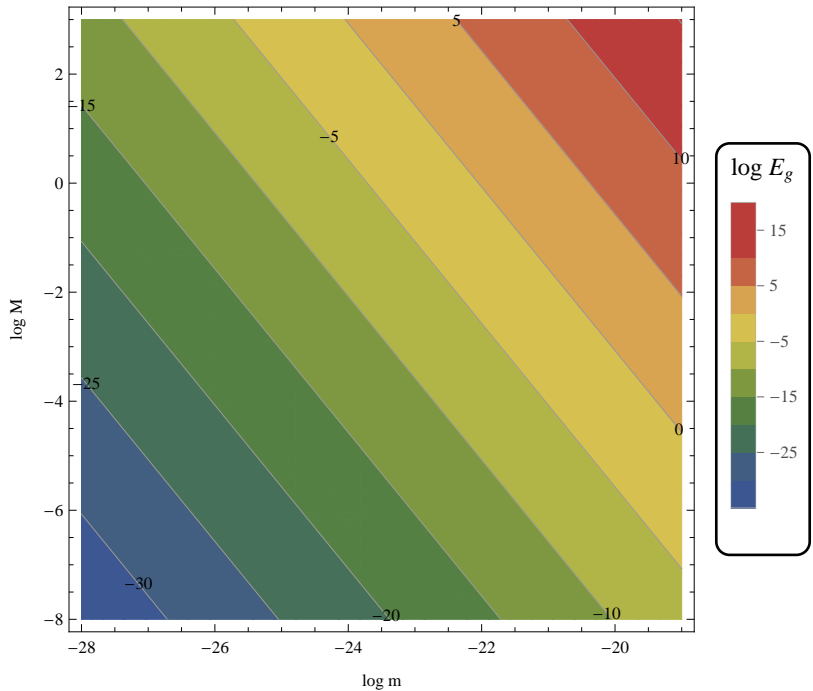}}
\end{center}
\caption{The quantum gravitational binding energy $E_g$ in eV, as a function of the quantum mechanical masses (``gravitational charges") $m$ and $M$, given in kg.}
\end{figure}

\begin{figure}
\begin{center}
\scalebox{1}{\includegraphics {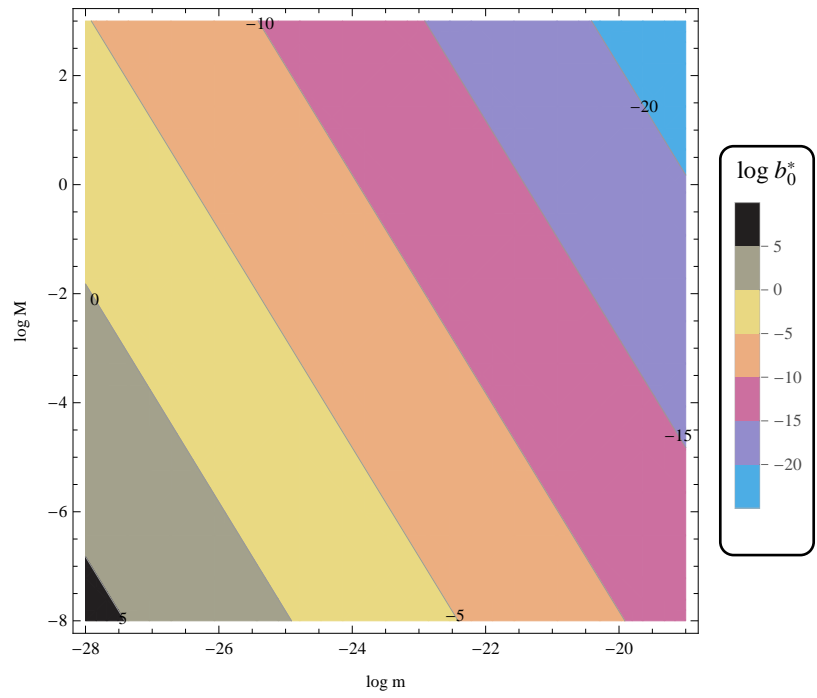}}
\end{center}
\caption{The ``gravitational Bohr-radius" $b^*_0$ in meters, as a function of the quantum mechanical masses $m$ and $M$, given in kg.}
\end{figure}

\begin{table}
  \begin{tabular}{|l|l|l|l|l|l|l|}
    \hline
    \multirow{2}{*}{$M$ (kg)} &
        \multicolumn{2}{c}{$m$ (kg) $10^{-20}$ \tiny{BEC}} &
      \multicolumn{2}{c}{$10^{-23}$ \tiny{BB}} &
        \multicolumn{2}{c|}{$10^{-27}$ \footnotesize{neutron}}
    \\
    & $E_g$ (eV) & $b_0^*$  (m) & $E_g$  (eV) & $b_0^*$ (m) & $E_g$ (eV) & $b_0^*$ (m)  \\
    \hline
     $10^{3}$  \, \tiny{SF} &  &  &  &  & $10^{-9}$ & $10^{-7}$ \\
    \hline
    $10^{-1}$ \tiny{SF}  & & & $10^{-5}$ & $10^{-11}$ &  &   \\
    \hline
    $10^{-2}$ \tiny{SF} & & &  $10^{-7}$ & $10^{-10}$ &  &  \\
    \hline
     $10^{-4}$ \tiny{SF} &   $10^{-2}$ & $10^{-14}$ &  & & & \\
    \hline
     $10^{-6}$ \tiny{SF} &  $10^{-6}$ & $10^{-12}$ &  & & &  \\
    \hline
  \end{tabular}
  \caption{Orders of magnitude for the quantum gravitational binding energy $E_g$ in eV, and the ``gravitational Bohr-radius" $b^*_0$ in meters, for a few potentially physically, \textit{i.e.} experimentally, interesting combinations of quantum masses $m$ and $M$, given in kg. SF = superfluid helium, BEC = gaseous Bose-Einstein condensate, BB = Buckyball ($C_{60}$) or similar ``macroscopic" quantum molecule. These are all known and well-studied objects in their own right. More speculatively (and outside the weak-field limit), an electron ($m \sim 10^{-30}$ kg) gravitationally bound to a Preon Star \cite{PS} with mass $M \sim 10^{12}$ kg tentatively gives [$E_g \sim 1$ eV, $b^*_0 \sim 10^{-10}$ m]; a neutrino ($m \sim 10^{-36}$ kg) bound to a Preon Star of $M \sim 10^{20}$ kg gives [$E_g \sim 10^{-2}$ eV, $b^*_0 \sim 10^{-6}$ m]. The characteristic size of a Preon Star is comparable to its Schwarzschild radius: $R_s (10^{12}$ kg) $\sim 10^{-15}$ m, $R_s (10^{20}$ kg) $\sim 10^{-7}$ m. The cosmic microwave background has an energy of $\sim 10^{-4}$ eV.}
\end{table}


\begin{thebibliography}{45}

\bibitem{Hansson} J.~Hansson, \textit{Aspects of nonrelativistic quantum gravity},
Braz.J.Phys. {\bf 39}, 707 (2009).
\href{http://arxiv.org/abs/0910.4289}{arXiv:0910.4289 [gr-qc]}

\bibitem{Feynman1} R.P.~Feynman,  p. 258 in \textit{The Role of Gravitation in Physics: Report from the 1957 Chapel Hill Conference},
Eds. C.M. DeWitt, D. Rickles, Edition Open Access (2011).

\bibitem{COW} R.~Colella, A.W.~Overhauser \& S.A.~Werner, \textit{Observation of Gravitationally Induced Quantum Interference},
Phys.Rev.Lett. {\bf 34}, 1472 (1975).

\bibitem{Nezvishevsky} V.V.~Nesvizhevsky \textit{et al.}, \textit{Quantum states of neutrons in the Earth's gravitational field}, Nature {\bf 415}, 297
(2002); \textit{Measurement of quantum states of neutrons in the Earth's gravitational field}, Phys.Rev. D {\bf 67}, 102002 (2003).

\bibitem{NS} A.~Lyne, F.~Graham-Smith,
{\it Pulsar Astronomy}, Third Edition (Cambridge University Press 2005).

\bibitem{Feynman} R.P.~Feynman, \textit{Application of Quantum Mechanics to Liquid Helium}, pp. 17-53 in Progress in Low Temperature Physics, Ed. by C.J. Gorter (North-Holland 1955).

\bibitem{PS} J.~Hansson, F.~Sandin,
{\it Preon stars: a new class of cosmic compact objects}, Phys. Lett.  {\bf B616}, 1 (2005).
\href{http://arxiv.org/abs/astro-ph/0410417}{arXiv:astro-ph/0410417}; F.~Sandin, J.~Hansson,
{\it Observational legacy of preon stars: Probing new physics beyond the CERN LHC}, Phys. Rev.  {\bf D76}: 125006 (2007).
\href{http://arxiv.org/abs/astro-ph/0701768}{arXiv:astro-ph/0701768.}

\end{thebibliography}
\end{document}